\documentclass[twocolumn,twoside,preprintnumbers,amsmath,amssymb,pacs]{revtex4}

\usepackage{epsfig}
\usepackage{graphicx}
\usepackage{dcolumn}
\usepackage{bm}

\usepackage{fancyhdr}

\usepackage{pslatex}

\begin{document}

\title{{\Large  The 2d Gross-Neveu Model at Finite Temperature and Density with Finite
$N$ Corrections }}

\author{Jean-Lo\"{\i}c Kneur}
\email{kneur@lpta.univ-montp2.fr}
\affiliation{Laboratoire de Physique  Th\'{e}orique et Astroparticules - CNRS
  - UMR 5207 Universit\'{e} Montpellier II, France}

\author{Marcus Benghi Pinto}
\email{marcus@fsc.ufsc.br}
\affiliation{Departamento de F\'{\i}sica, Universidade Federal de Santa
  Catarina, 88040-900 Florian\'{o}polis, Santa Catarina, Brazil}

\author{Rudnei O. Ramos}
\email{rudnei@uerj.br}
\affiliation{Departamento de
  F\'{\i}sica Te\'orica, Universidade do Estado do Rio de Janeiro, 20550-013
  Rio de Janeiro, RJ, Brazil}

\received{on 24 March, 2006}

\begin{abstract}



We use the linear $\delta$ expansion, or optimized perturbation theory, to evaluate the effective potential for the two dimensional Gross-Neveu model at finite temperature and density obtaining analytical equations for the critical temperature, chemical potential and fermionic mass which include finite $N$ corrections. Our results seem to improve over the traditional large-$N$ predictions.
\end{abstract}

\maketitle

\thispagestyle{fancy}
\setcounter{page}{0}

\section{Introduction}
The development of reliable analytical non-perturbative techniques to treat problems related to phase transitions in quantum chromodynamics (QCD) represents an important domain of research within quantum field theories. The appearance of large infrared divergences, happening for example in massless
field theories, like in QCD \cite{gross}, close to critical temperatures (in
field theories displaying a second order phase transition or a weakly first
order transition \cite{GR}) can only be dealt with in a non-perturbative fashion.   Among the analytical non-perturbative techniques one of the most used is   the $1/N$
approximation \cite{largeNreview}.  Though a powerful resummation method, this approximation can quickly become cumbersome after the resummation of the
first leading contributions, like for the $N=3$ case which regards QCD.
 This is due to technical difficulties such as the formal
resummation of infinite subsets of {}Feynman graphs and their subsequent
renormalization. In this work we  employ an alternative non-perturbative method known as the linear $\delta$ expansion (LDE) \cite {lde} to investigate the breaking and restoration of chiral symmetry within the two dimensional Gross-Neveu model \cite {gn}  at finite temperature ($T$) and chemical potential ($\mu$).
As we shall see, the LDE great advantage is that the actual selection and evaluation,including renormalization, of the relevant contributions are carried out in a completely perturbative way. Non-perturbative results are generated through the use of a variational optimization procedure known as the principle of minimal sensitivity (PMS) \cite {pms}.
The two dimensional Gross-Neveu model offers a perfect testing ground for the LDE-PMS because, apart from sharing common features with QCD, it is exactly solvable in the large-$N$ limit. The large-$N$ result for the critical temperature (at zero chemical potential) of the Gross-Neveu model is $T_c \simeq 0.567\,m_F(0)$ where $m_F(0)$ is the fermionic mass at $T=0$.  However, due to the appearance of kink--anti-kink configurations, the exact critical temperature for this model should be zero \cite {landau}. Because kink configurations are unsuppressed the system is segmented into regions of alternating signs of the order parameter, at low temperatures. Then, the net average value of the order parameter is zero. At leading order, the $1/N$ approximation misses this effect because the energy per kink goes to infinity as $N \to \infty$ while the contribution from the kinks has the form $e^{-N}$. Our strategy will be twofold. First, we show that the LDE-PMS exactly reproduces, within the $N \to \infty$ limit, the ``exact"  large-$N$ result. Next we show explicitely that already at the first non trivial order the LDE takes into account finite $N$ corrections which induce a lowering of $T_c$ as predicted by Landau's theorem. Here, the calculations are performed for three cases which are: (a) $T=0$ and $\mu=0$, (b) $T\ne 0$ and  $\mu=0$ and (c) $T=0$ and $\mu \ne 0$.  Our main results include analytical relations for the fermionic mass at $T=0$ and $\mu=0$, $T_c$ (at $\mu=0$) and $\mu_c$ (at $T=0$) which include finite $N$ corrections. The case $T \ne 0$ and $\mu \ne 0$, which allows for the determination of the tricritical points and phase diagram  is more complex, due to the numerics. This situation  is currently being treated by the present authors \cite {novogn}. In the next section we review the Gross-Neveu effective potential at finite temperature and chemical potential in the large-$N$ approximation. The LDE evaluations are presented in section III. The results are  discussed in section IV while section V contains our conclusions.

\section{The Gross-Neveu effective potential at finite temperature and chemical potential in the large-$N$ approximation}
The Gross-Neveu model is described by the Lagrangian density for a fermion
field $\psi_k$ ($k=1,\ldots,N$) given by \cite{gn}

\begin{equation}
{\cal L} =
\sum_{k=1}^N \left [\bar{\psi}_{k} \left( i \not\!\partial\right) \psi_{k} +
m_F {\bar \psi_k} \psi_k
+ \frac {g^2}{2} ({\bar \psi_k} \psi_k)^2\right ]\;.
\label{GN}
\end{equation}
 When $m_F=0$ the theory
is invariant under the discrete transformation

\begin{equation}
\psi \to \gamma_5 \psi \,\,\,,
\end{equation}
displaying a discrete chiral symmetry (CS). In addition, Eq. (\ref{GN}) has a
global $SU(N)$ flavor symmetry.

{}For the studies of the Gross-Neveu model  in the large-$N$ limit it is
convenient to define the four-fermion interaction as $g^2 N = \lambda$. Since
$g^2$ vanishes like $1/N$, we then study the theory in the large-$N$
limit with fixed $\lambda$ \cite {gn}. As usual, it is useful to
rewrite Eq. (\ref{GN}) expressing it in terms of an auxiliary (composite)
field $\sigma$, so that \cite {coleman}

\begin{equation}
{\cal L} =
\bar{\psi}_{k} \left( i \not\!\partial\right) \psi_{k}
-\sigma {\bar \psi_k} \psi_k
- \frac {\sigma^2 N}{2 \lambda}\;.
\label{Lsigma}
\end{equation}
As it is well known, using the $1/N$ approximation, the large-$N$ expression for
the effective potential is \cite {gn, coleman}

\begin{equation}
V_{\rm eff}^N(\sigma_c) = N \frac {\sigma_c^2}{2 \lambda} +
iN \int \frac {d^2 p}{(2\pi)^2}
\ln \left(p^2 - \sigma_c^2\right)\;.
\label{VN}
\end{equation}
The above equation can be extended at finite temperature and chemical potential applying the usual associations and replacements. E.g., momentum integrals of functions
$f(p_0,{\bf p})$  are replaced by
\begin{eqnarray}
\int  \frac {d^2 p}{(2 \pi)^2} f(p_0,{\bf p}) \to
iT \sum_n \int  \frac {d p}{(2 \pi)}
\; f[i(\omega_n  - i\mu),{\bf p}]\;,
\nonumber
\end{eqnarray}
where $\omega_n=(2 n +1) \pi T$, $n=0, \pm 1, \pm 2, \ldots$, are the
Matsubara frequencies for fermions \cite {kapusta}.  {}For the divergent, zero temperature
contributions, we choose dimensional regularization in arbitrary dimensions
$2\omega= 1-\epsilon$ and carry the renormalization in the $\overline{\rm
  MS}$ scheme, in which case the momentum integrals are written as

\[
\int \frac {dp}{(2 \pi)} \to \int_p = \left(\frac{e^{\gamma_E}
    M^2}{4 \pi} \right)^{\epsilon/2} \int \frac {d^{2\omega} p}{(2 \pi)^{2
    \omega}} \;,
\]

\noindent
where $M$ is an arbitrary mass scale and $\gamma_E \simeq 0.5772$ is the
Euler-Mascheroni constant. The integrals are then evaluated by using standard
methods.

In this case, Eq. (\ref{VN}) can be written as

\begin{equation}
\frac{V_{\rm eff}^N(\sigma_c)}{N} = \frac {\sigma_c^2}{2 \lambda} -
 T \sum_n \int  \frac {d p}{(2 \pi)}
\; \ln  \left[(\omega_n -i\mu)^2 + \omega^2_{p}(\sigma_c)
\right]\;,
\label{Veffn}
\end{equation}
where $\omega^2_{p}(\sigma_c) = {\bf p}^2 + \sigma_c^2$. The sum over the Matsubara's frequencies in Eq. (\ref{Veffn}) is also standard
\cite{kapusta} and gives for the effective potential, in the large-$N$
approximation, the result

\begin{eqnarray}
\frac {V_{\rm eff}^N(\sigma_c)}{N} &=&  \frac {\sigma_c^2}{2 \lambda} -
 \int_p
\omega_{p}(\sigma_c) \nonumber \\
&+& T \int_p  \ln\left(
1+ \exp\left\{-\left[\omega_{p}(\sigma_c)+\mu\right]/T\right\}
\right) \nonumber \\
&+&
T \int_p \ln\left( 1+\exp\left\{-\left[\omega_{p}(\sigma_c)-
\mu\right]/T\right\}
\right)  .
\label{Vefffull}
\end{eqnarray}
After integrating and renormalizing the above equation one obtains
\begin{equation}
\frac{V_{\rm eff}^N (\sigma_c)}{N} =
 \frac {\sigma_c^2}{2 \lambda} -
\frac{1}{2\pi} \left \{ \sigma_c^2 \left [ \frac {1}{2} +
\ln \left( \frac {M}{\sigma_c} \right ) \right] +
2 T^2 I_1(a,b) \right \} \;,
\label{VeffN}
\end{equation}
where
\begin{equation}
I_1(a,b) = \int_0^\infty dx  \left[ \ln \left( 1+e^{-\sqrt{x^2+a^2}-b} \right)
+ (b \to -b) \right]\;,
\label{Jab}
\end{equation}
with $a=\sigma_c/T$ and $b=\mu/T$.
Taking the $T=0$ and $\mu=0$ limit one may look for the effective potential minimum (${\bar \sigma}_c$) which, when different from zero signals dynamical chiral symmetry breaking (CSB). This minimization produces \cite {gn,coleman}
\begin{equation}
m_F(0)= {\bar \sigma}_c=M \exp\left(-\frac{\pi}{\lambda}\right).
\label{mF}
\end{equation}
\begin{figure}[htbp]
\begin{center}
\includegraphics[width=8cm]{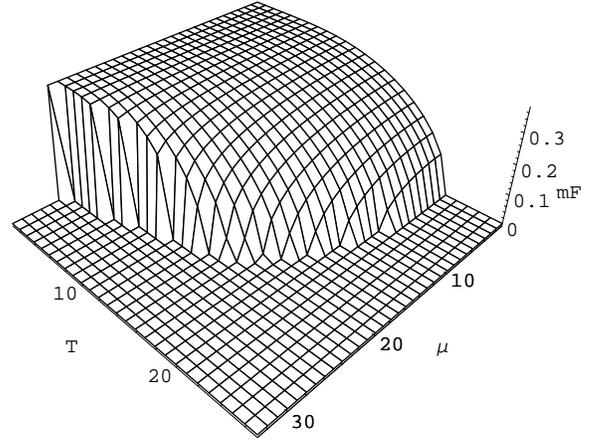}
\caption{A three dimensional graph showing the fermionic mass, $m_F$, as a function of $T$ and $\mu$. One sees a second order phase transition at $\mu=0$ while a first order transition occurs at $T=0$. All quantities are in units of $10 \times M$ while $\lambda=\pi$. }
\label{3d}
\end{center}
\end{figure}

\begin{figure}[htbp]
\begin{center}
\includegraphics[width=8cm]{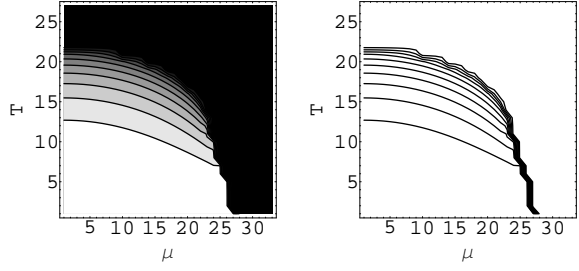}
\caption{Top views of figure 1. On the LHS we have a shaded figure where the black region represents CSR. The contour lines of the figure on the RHS indicate an abrupt (first order transition) for small values of $T$. Both figures display a (tricritical) point where the smooth descent meets the abrupt one.  All quantities are in units of $10 \times M$ while $\lambda=\pi$. }
\label{contour}
\end{center}
\end{figure}

One may proceed by numerically investigating $m_F$ as a function of
$T$ and $\mu$ as shown in Figure
\ref {3d} which shows a smooth phase (second order) transition at $\mu=0$. At this point,  the exact value for the critical temperature ($T_c$) at which chiral symmetry restoration (CSR) occurs  can be  evaluated analytically  producing \cite {wrongtc}
\begin{equation}
T_c = m_F(0) \frac{e^{\gamma_E}}{\pi} \simeq
0.567 \; m_F (0),
\label{Tc}
\end{equation}
while, according to  Landau's theorem, the exact result should be $T_c=0$.  By looking at Figure \ref {3d} one notices an abrupt (first order) transition when $T=0$. The analytical value at which this transition occurs has also been evaluated, in the large-$N$ limit, yielding \cite {muc}
\begin{equation}
\mu_c =\frac{ m_F(0)}{\sqrt{2}} \,.
\label{largeNmuc}
\end{equation}
In the $T-\mu$ plane there is a (tricritical) point where the lines describing the first and second order transition meet. This can be seen more clearly  by analyzing the top views of figure \ref {3d}. Figure \ref {contour} shows these top views in a way which uses shades (LHS figure) and contour lines (RHS figure). The tricritical point ($P_{tc}$) values can be numerically determined producing
$P_{tc}=(T_{tc},\mu_{tc})=[0.318\,m_F(0), 0.608\,m_F(0)]$ \cite {italianos}.

\section{The Linear $\delta$ Expansion and finite $N$ corrections to the effective potential}
 According to the usual LDE interpolation prescription \cite {lde}
 the deformed {\it original} four fermion theory displaying CS
reads

\begin{equation}
{\cal L}_{\delta} = \sum_{k=1}^N \left [
\bar{\psi}_{k} \left( i \not\!\partial\right) \psi_{k} +
\eta (1-\delta) {\bar \psi_k} \psi_k
+ \delta \frac {\lambda}{2N} ({\bar \psi_k} \psi_k)^2 \right ]\;.
\label{GNlde}
\end{equation}

\noindent
So, that at $\delta=0$ we have a theory of free fermions.  Now, the
introduction of an auxiliary scalar field $\sigma$ can be achieved by adding
the quadratic term,

\begin{equation}
- \frac{ \delta N}{2 \lambda} \left ( \sigma +
\frac {\lambda}{N} {\bar \psi_k} \psi_k \right )^2 \,,
\end{equation}
to ${\cal L}_{\delta}(\psi, {\bar \psi})$. This leads to the
interpolated model

\begin{equation}
{\cal L}_{\delta} = \sum_{k=1}^N \left [
\bar{\psi}_{k} \left( i \not\!\partial\right) \psi_{k} -
\delta \eta_* {\bar \psi_k} \psi_k
- \frac {\delta N }{2 \lambda } \sigma^2 + {\cal L}_{ct,\delta} \right ]  \;,
\label{GNdelta}
\end{equation}

\noindent
where $\eta_*= \eta -(\eta - \sigma_c)\delta$.
The counterterm Lagrangian density, ${\cal L}_{ct,\delta}$, has the same
polynomial form as in the original theory while the
coefficients are allowed to be $\delta$ and $\eta$ dependent. Details about renormalization within the LDE can be found in Ref. \cite{prd1}.

{}From the Lagrangian density in the interpolated form, Eq.  (\ref{GNdelta}),
we can immediately read the corresponding new Feynman rules in Minkowski space. Each Yukawa vertex carries a factor $-i \delta$ while the (free) $\sigma$ propagator is now $-i
\lambda/(N \delta)$. The  LDE dressed fermion propagator is

\begin{equation}
S_F(p)=\frac{i}{\not \! p - \eta_*+i\epsilon}\;,
\label{SF}
\end{equation}
where $\eta_*= \eta -(\eta - \sigma_c)\delta$.

\bigskip
\bigskip

\begin{figure}[htbp]
\begin{center}
\includegraphics[width=8cm]{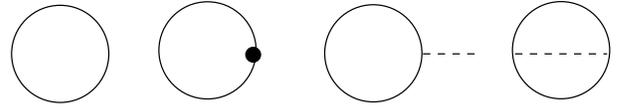}
\caption{LDE Feynman graphs contributing up to order-$\delta$. The black dot represents a $\delta \eta$ insertion. The external dashed line represents $\sigma_c$ while the internal line is the $\sigma$ propagator. The last diagram brings the first finite $N$ correction to the effective potential. }
\label{irqcd1}
\end{center}
\end{figure}
\bigskip

{}Finally, by summing up the contributions shown in figure \ref {irqcd1} one obtains the complete LDE expression to order-$\delta$

\begin{eqnarray}
\frac{V_{{\rm eff},\delta^1}}{N} (\eta) &=&
\delta \frac {\sigma_c^2}{2 \lambda} -
 \frac{1}{2\pi} \left \{ \eta^2 \left [ \frac {1}{2} + \ln \left (
\frac {M}{\eta} \right ) \right ] + 2 T^2 I_1(a,b) \right \}
\nonumber \\
&+& \delta \frac{\eta(\eta-\sigma_c)}{\pi}
\left[\ln\left(\frac{M}{\eta}\right) - I_2 (a,b) \right]
\nonumber \\
&+& \frac {\delta\lambda \eta^2}{4\pi^2\:N} \left \{
  \left [ \ln \left ( \frac {M}{\eta}  \right ) - I_2(a,b)
\right ]^2 +  J^2_2(a,b) \right \}\;.
\label{Vdelta1}
\end{eqnarray}

where $I_1(a,b)$ is defined by Eq. (\ref{Jab}), with  $a=\eta/T$. Also,

\begin{equation}
I_2(a,b)= \int_0^\infty \frac {d x}{\sqrt{x^2 +a^2}} \left(
\frac{1}{e^{\sqrt{x^2 + a^2} + b}+1}+ (b\to -b) \right) \;,
\label{I2ab}
\end{equation}
and
\begin{eqnarray}
J_2(a,b)
= \frac{\sinh(b)}{a} \int_0^\infty d x
\frac{1}{\cosh(\sqrt{x^2 + a^2})+\cosh(b)}  \;.
\label{J2ab}
\end{eqnarray}

\noindent
Notice once more, from Eq. (\ref{Vdelta1}), that our first order already takes into account finite $N$ corrections.  Now, one must fix the two non original parameters, $\delta$ and $\eta$, which appear in Eq. (\ref {Vdelta1}). Recalling that at $\delta=1$ one retrieves the original Gross-Neveu Lagrangian allows us to choose the unity as the  value for the dummy parameter $\delta$. The infra red regulator $\eta$ can be fixed by demanding $V_{{\rm eff}, \delta^1}$ to be evaluated at the point where it is less sensitive to variations with respect to $\eta$. This criterion, known as Principle of the Minimal Sensitivity (PMS) \cite {pms} can be written as
\begin{equation}
\frac {d V_{{\rm eff}, \delta^1}}{d \eta}\Big |_{\bar \eta, \delta=1} = 0 \;.
\label{PMS}
\end{equation}
In the next section the PMS will be used to generate the non-perturbative optimized LDE results.
\section{ Optimized Results}
{}From the PMS procedure we then obtain from Eq. (\ref
{Vdelta1}), at $\eta = {\bar \eta}$, the general result
\begin{widetext}
\begin{equation}
\left \{ \left [ {\cal Y}(\eta,T,\mu)
+ \eta \frac {d}{d \eta} {\cal Y}(\eta,T,\mu) \right ]
\left[ \eta - \sigma_c +
\eta \frac{\lambda}{2 \pi N} {\cal Y}(\eta,T,\mu) \right]
+ \frac{\lambda T^2}{2 \pi N}J_2(\eta/T,\mu/T)
\frac {d}{d \eta}J_2(\eta/T,\mu/T)
\right\}\Bigr|_{\eta = \bar{\eta}} = 0 \;,
\label{genpms}
\end{equation}
\end{widetext}
where we have defined the function

\begin{equation}
{\cal Y}(\eta,T,\mu) =\ln \left ( \frac {M}{\eta} \right ) -
I_2(\eta/T,\mu/T) \;.
\end{equation}
Let us first consider the case $N \to \infty$. Then, Eq. (\ref{genpms}) gives two solutions where the first one is $\bar \eta = \sigma_c$ which, when plugged in Eq. (\ref {Vdelta1}), exactly reproduces the large-$N$ effective potential, Eq. (\ref {VeffN}). This result was shown to rigorously hold at any order in $\delta$ provided that one stays within the large-$N$ limit \cite {npb}. The other possible solution, which depends only upon the scales $M$,$T$ and $\mu$, is considered unphysical \cite {npb}.

\subsection{The case $T=0$ and $\mu=0$}

Taking Eq. (\ref {genpms}) at $T=\mu=0$  one gets

\begin{equation}
\left [ \ln \left ( \frac {M}{{\bar \eta}} \right ) - 1 \right ]
\left [ {\bar \eta} - \sigma_c - {\bar \eta} \frac {\lambda}{2 \pi N}
\ln \left ( \frac {{\bar \eta}}{M} \right ) \right ]=0 \,\,.
\label{pmstmuzero}
\end{equation}

\noindent
As discussed previously, the first factor leads to the model independent
result, ${\bar \eta} = M/e$, which we shall neglect.  At the same time the
second factor in (\ref{pmstmuzero}) leads to a self-consistent gap equation
for $\bar \eta$, given by

\begin{equation}
{\bar \eta}_{\delta^1} (\sigma_c)= \sigma_c \left[ 1- \frac {\lambda}{2 \pi N}
 \ln \left ( \frac {{\bar \eta}_{\delta^1}}{M} \right )\right]^{-1} \;.
\label{etatzero}
\end{equation}

\noindent
The solution for $\bar \eta_{\delta^1}$ obtained from Eq. (\ref{etatzero}) is

\begin{equation}
\bar{\eta}_{\delta^1}(\sigma_c) = M \exp\left\{ \frac{2 \pi N}{\lambda} +
W\left[-\frac{2\pi N}{\lambda} \frac{\sigma_c}{M}\, \exp\left(
-\frac{2 \pi N}{\lambda} \right)\right]  \right\}  \;,
\label{etabarsol}
\end{equation}
where $W(x)$ is the Lambert $W$ function, which satisfies $W(x)
\exp[W(x)] = x$.

To analyze CS breaking we then replace $\eta$ by Eq. (\ref{etabarsol}) in
Eq. (\ref {Vdelta1}), which is taken at $T=0$ and $\mu=0$.  As usual, CS breaking
appears when the effective potential displays minima at some particular value
${\bar \sigma_c} \ne 0$. Then, one has to solve

\begin{equation}
\frac{V_{{\rm eff},\delta^1}(\sigma_c,\eta=\bar{\eta}_{\delta^1})}{d \sigma_c}
\Bigr|_{\delta=1,\sigma_c=\bar{\sigma}_c} =0\;.
\label{dVeffeta}
\end{equation}
Since $m_F=\bar{\sigma}_c$, after some algebraic manipulation of Eq.
(\ref{dVeffeta}) and using the properties of the $W(x)$ function, one finds

\begin{equation}
m_{F} (T=0,\mu=0) =
M {\cal F} (\lambda,N)\left( 1 - \frac {1}{2N} \right)^{-1}\;,
\label{allmf}
\end{equation}
where we have defined the quantity ${\cal F}(\lambda,N)$ as

\begin{equation}
{\cal F}(\lambda,N)= \exp \left \{ -\frac {\pi}{\lambda[1-1/(2N)]}
\right \}\;.
\label{funcF}
\end{equation}
\begin{figure}[htbp]
\begin{center}
\includegraphics[width=8cm]{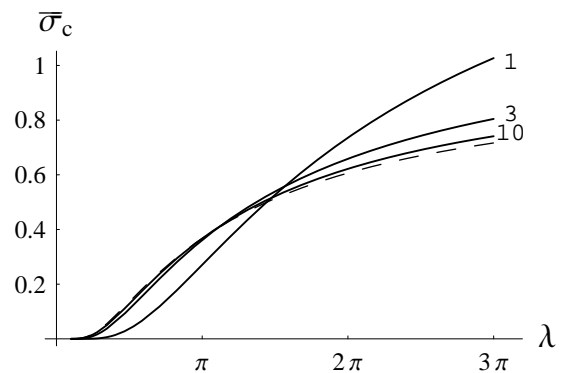}
\caption{The effective potential minimum,${\bar \sigma}_c$, which corresponds to the  fermionic mass, as a function of $\lambda$ for $N=1,3$ and $10$. The dashed line represents the large-$N$ result. ${\bar \sigma}_c$ is given in units of $M$.}
\label{alln}
\end{center}
\end{figure}
Eq. (\ref{allmf}) is our result for the fermionic mass at first order in
$\delta$ which goes beyond the large-$N$ result, Eq. (\ref{mF}). Note  that
in the $N \to \infty$ limit, ${\cal F}(\lambda,N \to \infty) = \exp ( -
\pi/\lambda)$. Therefore, Eq. (\ref{allmf}) correctly reproduces, within the LDE non
perturbative resummation, the large-$N$ result, as already discussed.
In {}Fig. \ref{alln} we compare the order-$\delta$ LDE-PMS results for
$\bar{\sigma}_c$ with the one provided by the large-$N$ approximation.
One can now obtain an analytical result for ${\bar \eta}$ evaluated at ${\bar \sigma}_c=\sigma_c$. Eqs. (\ref{etabarsol}) and
(\ref{allmf}) yield
\begin{equation}
\bar{\eta}_{\delta^1}(\bar{\sigma}_c) = M {\cal F}(\lambda,N)\;.
\label{alleta}
\end{equation}
Fig.  \ref{allguessfig} shows that $\bar{\eta}({\bar \sigma}_c)$
is an increasing function of both $N$ and $\lambda$ kickly saturating for $N \gtrsim 3$. The same figure shows the results obtained numerically with the PMS.

\begin{figure}[htbp]
\begin{center}
\includegraphics[width=8cm]{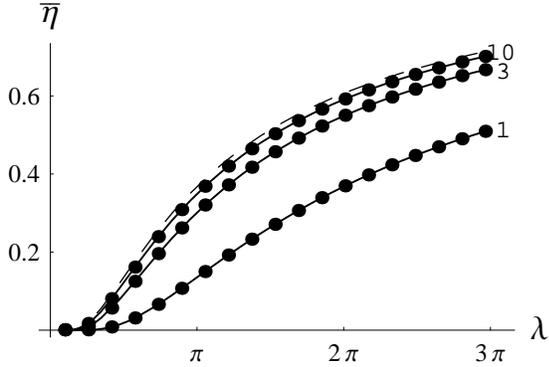}
\caption{The LDE optimum mass (${\bar \eta}$), evaluated at $\sigma_c = {\bar \sigma}_c$, as a function of $\lambda$ for $N=1,3$ and $10$. The continuous lines were obtained from the analytical result, Eq. (\ref {alleta}), while the dots represent the results of numerical optimization. ${\bar \eta}$ is given in units of $M$.}
\label{allguessfig}
\end{center}
\end{figure}

\subsection{The case $T \ne 0$ and $\mu=0$}

Let us now investigate the case $T \ne 0$ and $\mu=0$. In principle, this could be done numerically by a direct application of the PMS the LDE effective potential, Eq. (\ref{Vdelta1}). However, as we shall see, neat analytical results can be obtained if one uses the high temperature expansion by taking
 $\eta/T = a \ll 1$ and $\mu/T = b \ll 1$. The validity of such action could be questioned, at first, since $\eta$ is arbitrary. However, we have cross checked the PMS results obtained analytically using the high $T$ expansion with the ones obtained numerically without using this approximation. This cross check shows a good agreement between both results. Expanding Eq.
(\ref{Jab}) in powers of $a$ and $b$, the result is finite and given by
\cite{zhou}

\begin{eqnarray}
I_1 (a\ll 1,b\ll 1) &=& \frac{\pi^2}{6} + \frac{b^2}{2} - \frac{a^2}{2}
\ln \left(\frac{\pi}{a} \right) - \frac{a^2}{4}(1-2\gamma_E)
\nonumber \\
&-& \frac{7 \zeta(3)}{8 \pi^2} a^2 \left(b^2 + \frac{a^2}{4} \right)
+{\cal O}(a^2 b^4, a^4 b^2)\;, \nonumber\\
\label{JT}
\end{eqnarray}
and
\begin{equation}
I_2(a,b) = \ln\left(\frac{\pi}{a}\right) - \gamma_E
+\frac{7 \xi(3)}{4 \pi^2} \left(b^2+\frac{a^2}{2}\right) +
{\cal O}(a^4,b^4)\;,
\label{highTI2}
\end{equation}
where $\zeta(3) \simeq 1.202$.
 If we then expand
Eq. (\ref{Vdelta1}) at high temperatures, up to order $\eta^2/T^2$, we obtain

\begin{eqnarray}
\frac{V_{{\rm eff},\delta^1}}{N} &=&
\delta \frac {\sigma_c^2}{2 \lambda} - T^2 \frac {\pi}{6} -
 \frac{\eta^2}{2\pi}  \left [ \ln \left (
\frac {M e^{\gamma_E}}{T \pi} \right )  - \frac{ 7 \zeta(3)}{4 (2\pi)^2}
\frac{\eta^2}{T^2} \right ]
\nonumber \\
&+& \delta \frac{\eta(\eta-\sigma_c)}{\pi} \left [ \ln \left (
\frac {M e^{\gamma_E}}{T \pi} \right )  - \frac{ 7 \zeta(3)}{2 (2\pi)^2}
\frac {\eta^2}
{T^2} \right ]
\nonumber \\
&+& \frac {\delta \lambda \eta^2 }{(2\pi)^2N}
 \left [ \ln^2 \left (
\frac {M e^{\gamma_E}}{T \pi} \right )  - \frac{ 7 \zeta(3)}{ (2\pi)^2}
\ln \left (
\frac {M e^{\gamma_E}}{T \pi} \right ) \frac{\eta^2}{T^2}   \right ] \,.\nonumber\\
\label{Vdelta1hit}
\end{eqnarray}
Now, one sets $\delta=1$ and applies the PMS to Eq. (\ref {Vdelta1hit}) to obtain the optimum LDE mass

\begin{eqnarray}
{\bar \eta}(\sigma_c,T) &=& \sigma_c \left \{ 1+ \frac {\lambda}{N(2\pi)}
\left [ \ln \left (
\frac {M e^{\gamma_E}}{T \pi} \right ) \right . \right . \nonumber \\
& -& \left . \left . \frac{ 7 \zeta(3)}{2 (2\pi)^2} \frac
{\sigma_c^2}
{T^2} \left [ 1+ \frac {\lambda}{N(2\pi)}  \ln \left (
\frac {M e^{\gamma_E}}{T \pi} \right ) \right ]^{-2}
 \right ] \right \}^{-1} \,\,. \nonumber\\
\label {etafinitet}
\end{eqnarray}

The above result is plugged back into Eq. (\ref {Vdelta1hit}) which, for consistency, should be re expanded to the order $\eta^2/T^2$. This generates a nice analytical result for the thermal fermionic mass
\begin{eqnarray}
{\bar \sigma_c}(T) &=& \pm \frac {T}{N^2 \sqrt{14 \pi \zeta(3)\lambda}}
\left [ 2N \pi + \ln \left (
\frac {M e^{\gamma_E}}{T \pi} \right ) \right ]^{3/2} \nonumber \\
&\times&\left [ - 2N\pi+(2N-1) \lambda \ln \left (
\frac {M e^{\gamma_E}}{T \pi} \right ) \right ]^{1/2}\;.
\label{sigmalde}
\end{eqnarray}

\begin{figure}[htbp]
\begin{center}
\includegraphics[width=8cm]{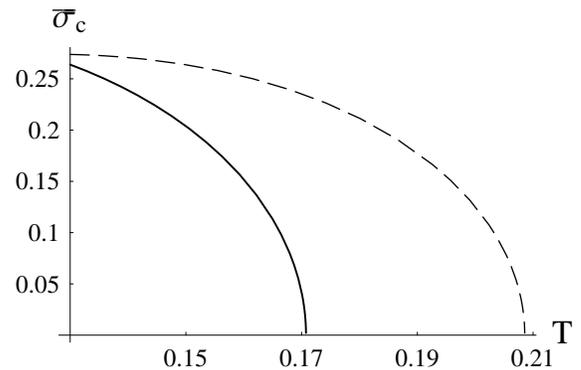}
\caption{The effective potential minimum, ${\bar \sigma}_c$, as a function of the temperature. Both quantities are in units of $M$ and have been plotted for $N=3$ and $\lambda=\pi$. The dotted line corresponds to the large result predicting $T_c=0.208\,M$ while the continuous line, which represents the LDE result, predicts $T_c=0.170\,M$. In both cases the transition is of the second kind.}
\label{mFhighT}
\end{center}
\end{figure}

{}Figure \ref{mFhighT} shows ${\bar \sigma_c}(T)/M$ given by Eq. (\ref {sigmalde}) as a function of $T/M$, again
showing a continuous (second order) phase transition for CS
breaking/restoration.
The numerical results illustrated by {}Fig. \ref{mFhighT}
show that the transition is of the second kind and an analytical equation for
the critical temperature can be obtained by requiring that the minima vanish at
$T_c$. {}From Eq. (\ref {sigmalde}) one sees that ${\bar \sigma_c}(T=T_c)=0$
can lead to two possible solutions for $T_c$.

\begin{figure}[htbp]
\begin{center}
\includegraphics[width=8cm]{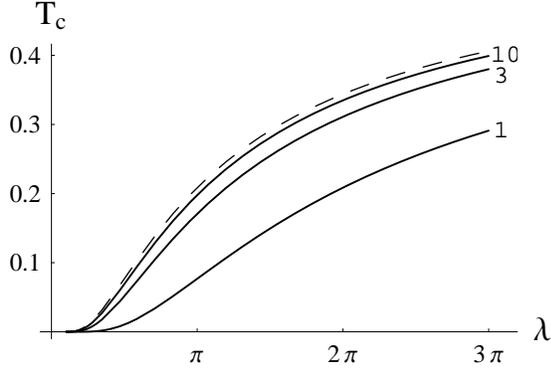}
\caption{The critical temperature ($T_c$), in units of $M$, as a function of $\lambda$ for $N=1,3$ and $10$. The continuous lines represent the LDE results while the dotted line represents the large-$N$ result.}
\label{allTc}
\end{center}
\end{figure}
\bigskip
 The one coming from
\begin{equation}
\left [ 2N \pi + \ln \left (
\frac {M e^{\gamma_E}}{T_c \pi} \right ) \right ] = 0\;,
\end{equation}

\noindent
can easily be seen as not been able to reproduce the known large-$N$ result,
when $N\to \infty$, $T_c = M \exp(\gamma_E-\pi/\lambda)/\pi$.  However, the
other possible solution coming from

\begin{equation}
\left [ - 2N\pi+(2N-1) \lambda \ln \left (
\frac {M e^{\gamma_E}}{\pi T_c } \right ) \right ] =0\;,
\end{equation}
gives for the critical temperature, evaluated at first order in $\delta$, the
result

\begin{equation}
T_{c} = M \frac{e^{\gamma_E}}{\pi}   \exp
\left \{ - \frac {\pi}{\lambda[1 -1/(2N)]} \right \} \,
=\, M\frac{e^{\gamma_E}}{\pi} {\cal F}(\lambda,N) \,\, \;,
\label{ldetc}
\end{equation}
with ${\cal F}_\lambda (N)$ as given before, by Eq. (\ref{funcF}).  Therefore, Eq.
(\ref{ldetc}) also exactly reproduces the large-$N$ result for $N \to \infty$.  The
results given by this equation are plotted in {}Fig.  \ref{allTc} in terms of
$\lambda$ for different values of $N$. The (non-perturbative) LDE results show that $T_c$ is always
smaller (for the realistic finite $N$ case) than the value
predicted by the large-$N$ approximation. According to Landau's theorem for
phase transitions in one space dimensions,  our LDE
results, including the first $1/N$ correction, seem to converge to the
right direction.

\subsection{The case $T = 0$ and $\mu \ne 0$}

One can now study the case $T=0,\mu\ne0$ by taking the limit $T\to 0$ in the integrals $I_1$, $I_2$ and $J_2$ which appear in the LDE effective potential, Eq. (\ref {Vdelta1}). In this limit, both functions are given by
\begin{eqnarray}
&& \lim_{T\to 0} T^2 I_1(a,b)=
- \frac{1}{2} \theta(\mu - \eta)
\left[ \eta^2 \ln \left( \frac{\mu + \sqrt{\mu^2 -
\eta^2}}{\eta}
\right) \right . \nonumber \\
&-& \left . \mu \sqrt{\mu^2 - \eta^2} \right]\;,
\label{J2T0}
\\
&&
\lim_{T \to 0} I_2(a,b) =  \theta(\mu-\eta)
\ln\left(\frac{\mu +\sqrt{\mu^2 - \eta^2}}{\eta} \right)\;,
\label{I2T0}
\\
&&
\lim_{T \to 0} T J_2(a,b) =  {\rm sgn}(\mu) \theta(\mu-\eta)
\sqrt{\mu^2 - \eta^2} \;.
\label{JJ2T0}
\end{eqnarray}

\begin{figure}[htbp]
\begin{center}
\includegraphics[width=8cm]{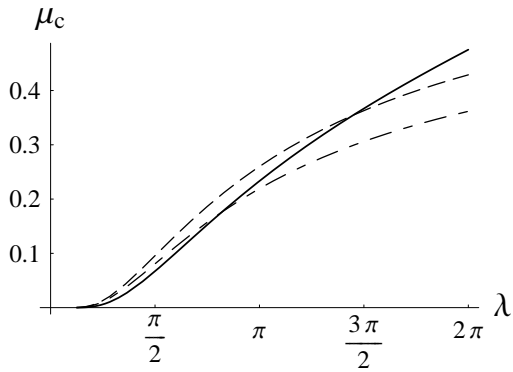}
\caption[]{\label{gnfig13} The critical chemical potential $\mu_c$ in units
  of $M$, plotted as a function of $\lambda$ for $N=3$ and $T=0$. The dashed
  line represents the $1/N$ result at leading order, the dot-dashed line
  represents the $1/N$ result at next to leading order and the continuous line
  is the first order LDE result.}
\label{fig9}
\end{center}
\end{figure}

Then, one has to analyze two situations.
In the first, $\eta > \mu$, the optimized $\bar \eta$
is given by

\begin{eqnarray}
&&\left \{ \left [ \ln \left ( \frac{M}{\eta} \right ) - 1 \right ]
\left [ {\eta} - \sigma_c + \frac {\lambda \eta}{2 \pi N}
\ln \left ( \frac{M}{\eta} \right ) \right ] \right . \nonumber \\
&-& \left .\frac {\lambda \mu^2}{2 \pi N} \frac {1}{{\eta}}
\ln \left ( \frac{M}{\eta} \right ) \right \}\Bigr|_{\eta=\bar \eta} = 0 \,,
\label{pmstmu}
\end{eqnarray}
while for the second, $\eta < \mu$, $\bar{\eta}$ is found from the solution of
\begin{widetext}
\begin{equation}
\left\{
\left[ \eta-\sigma_c-\frac{\lambda \eta}{2 \pi N}
\ln  \left( {\frac {\mu+\sqrt {{\mu}^{2}-{\eta}^{2}}}{M}} \right) \right]
\left[ - \ln  \left( {\frac {\mu+\sqrt {{\mu}^{2}-{\eta}
^{2}}}{M}} \right) - \frac{\eta^2}{(\eta^2 -\mu^2 - \mu \sqrt {\mu^2- \eta^2})} \right]
- \frac{\lambda \eta }{2 \pi N}
\right\}
\Bigr|_{\eta=\bar \eta}  =0\;.
\label{mupms}
\end{equation}
\end{widetext}
Note that the results given by  Eqs. (\ref{J2T0}-\ref{JJ2T0})  vanish for $\mu < \eta$.
{}Fig. \ref{fig9}  shows  $\mu_c$, obtained numerically, as a function of $\lambda$ for
different values of $N$. Our result is contrasted with the ones furnished by the $1/N$ approximation. The analytical expressions for
$\bar{\eta}_{\delta^1}({\bar \sigma}_c)$, Eq. (\ref{alleta}), and $T_{c}$, Eq.
(\ref{ldetc}), suggest that an approximate solution for $\mu_{c}$ at first order
in $\delta$ is given by

\begin{equation}
\mu_{c} (T=0) \simeq \frac {M}{\sqrt 2} {\cal F}(\lambda,N)\;.
\label{allmu}
\end{equation}

It is interesting to note that both results,
for $T_{c}$, Eq. (\ref{allTc}), and $\mu_{c}$, Eq.
(\ref{allmu}), follow exactly the same trend as the corresponding results obtained from
the large-$N$ expansion, Eqs. (\ref{Tc}) and (\ref{largeNmuc}), respectively,
which have a common scale given by the zero temperature and density fermion
mass $m_F(0)$. Here, the common scale is given by $\bar{\eta}$ evaluated
at $\sigma_c = {\bar \sigma}_c$ and $T=\mu=0$, $\bar{\eta}_{\delta^1}({\bar \sigma}_c)= M {\cal F}(\lambda,N)$.
\section{Conclusions}

We have used the non-perturbative linear $\delta$ expansion method (LDE) to evaluate the effective potential of the two dimensional Gross-Neveu model at finite temperature and chemical potential. Our results show that when one stays within the large-$N$ limit the LDE correctly reproduces the $1/N$ approximation leading order results for the fermionic mass, $T_c$ and $\mu_c$. However, as far as $T_c$ is concerned the large-$N$ predicts $T_c \simeq 0.567\, m_F(0)$ while Landau's theorem for phase transitions in one space dimensions  predicts $T_c=0$. Having this in mind we have considered the first finite $N$ correction to the LDE effective potential. The whole calculation was performed with the easiness allowed by perturbation theory. Then, the effective potential was optimized in order to produce the desired non-perturbative results. This procedure has generated analytical relations for the relevant quantities (fermionic mass, $T_c$ and $\mu_c$) which explicitely display finite $N$ corrections. The relation for $T_c$, for instance, predicts smaller values than the ones predicted by the large-$N$ approximation which hints on the good convergence properties of the LDE in this case. The LDE convergence properties in critical temperatures has received support by recent investigations concerned with the evaluation of the critical temperature for weakly interacting homogeneous Bose gases \cite {prl}. In order to produce the complete phase diagram, including the tricritical points, we are currently investigating the case $T \ne 0$ and  $\mu \ne 0$ \cite {novogn}.

\acknowledgments

M.B.P. and R.O.R. are partially supported by CNPq. R.O.R. acknowledges partial support from FAPERJ and M.B.P.  thanks the organizers of IRQCD06  for the invitation.



\begin{thebibliography}{99}
\bibitem{gross} D. J. Gross, R. D. Pisarski and L. G. Yaffe, Rev. Mod. Phys.
  {\bf 53}, 43 (1981).

\bibitem{GR}M. Gleiser and R. O. Ramos, Phys. Lett. {\bf B300}, 271 (1993); J. R. Espinosa, M. Quir\'os and F. Zwirner, Phys. Lett.  {\bf
    B291}, 115 (1992).

\bibitem{largeNreview}M. Moshe and J. Zinn-Justin, Phys. Rept. {\bf 385}, 69
  (2003).

\bibitem{lde}  A. Okopinska, Phys. Rev.  {\bf D35}, 1835 (1987);
A. Duncan and M. Moshe, Phys. Lett. {\bf B215}, 352 (1988).

\bibitem{gn} D. Gross and A. Neveu, Phys. Rev. {\bf D10}, 3235 (1974).

\bibitem{pms} P. M. Stevenson, Phys. Rev. {\bf D23}, 2961 (1981); Nucl. Phys.
  {\bf B203}, 472 (1982).

\bibitem {landau} L.D. Landau and E.M. Lifshtiz, {\it Statistical Physics}
  (Pergamon, N.Y., 1958) p. 482; R.F. Dashen, S.-K. Ma and R. Rajaraman, Phys. Rev. {\bf D11}, 1499 (1974); S.H. Park, B. Rosenstein and B. Warr, Phys. Rept. {\bf 205}, 59 (1991).


\bibitem{novogn} J.-L. Kneur, M.B. Pinto and R.O. Ramos, Phys. Rev. {\bf D74}, 125020 (2006).

\bibitem{coleman} S. Coleman, \textit{Aspects of Symmetry} (Cambridge
University Press, Cambridge, 1985).

\bibitem{kapusta}J. I. Kapusta, \textit{Finite-Temperature Field Theory}
(Cambridge University Press, Cambridge, England, 1985).

\bibitem{wrongtc} L. Jacobs, Phys. Rev {\bf D10}, 3976 (1974); B.J.
  Harrington and A. Yildz, Phys. Rev. {\bf D11}, 779 (1974).

\bibitem{muc} U. Wolff, Phys. Lett. {\bf B157}, 303 (1985); T.F. Treml, Phys. Rev. {\bf D39}, 679 (1989).

\bibitem{italianos} A. Barducci, R. Casalbuoni, M. Modugno and G. Pettini,
  Phys. Rev. {\bf D51}, 3042 (1995).

\bibitem{prd1} M. B. Pinto and R. O. Ramos, Phys. Rev. {\bf D60}, 105005
  (1999); {\it ibid.} {\bf D61}, 125016 (2000); J.-L. Kneur and D. Reynaud, JHEP {\bf 301},14 (2003).

\bibitem{npb} S.K. Gandhi, H.F. Jones and M.B. Pinto, Nucl. Phys. {\bf B359},
  429 (1991).

\bibitem{zhou} B. R. Zhou, Phys. Rev. {\bf D57}, 3171 (1998); Comm. Theor.
  Phys. {\bf 32}, 425 (1999).

\bibitem{prl}J.-L. Kneur, M. B. Pinto and R. O. Ramos, Phys. Rev. Lett.  {\bf
    89}, 210403 (2002); Phys. Rev. {\bf A68}, 043615 (2003).

\end{thebibliography}
\end{document}